\documentclass[conference]{IEEEtran}

\usepackage{amsmath, amssymb, amsfonts}

\usepackage{algorithm}
\usepackage{algorithmic}

\usepackage{array}
\usepackage{booktabs}
\usepackage{multirow}

\usepackage{graphicx}
\usepackage{float}
\usepackage{stfloats}
\usepackage[section]{placeins}

\usepackage{textcomp}
\usepackage{url}
\usepackage{verbatim}
\usepackage{cite}
\usepackage{microtype}    
\usepackage{cuted}        
\usepackage{balance}      

\usepackage{xcolor}
\definecolor{myteal}{HTML}{008080}

\usepackage{hyperref}
\hypersetup{
    colorlinks=true,
    urlcolor=blue,
    citecolor=blue,
}

\newcommand\blfootnote[1]{%
  \begingroup
  \renewcommand\thefootnote{}\footnote{#1}%
  \addtocounter{footnote}{-1}%
  \endgroup
}
\usepackage{tikz}
\usetikzlibrary{shapes.geometric, arrows.meta, positioning, calc}

\begin{document}

\title{Fast Diffusion with Physics-Correction for ACOPF}

\author{
Shashank Shekhar$^\star$,
Abhinav Karn$^\star$,
Kris Keshav$^\star$,
Shivam Bansal$^\star$,
and Parikshit Pareek$^\dagger$\\
Department of Electrical Engineering, Indian Institute of Technology Roorkee (IIT Roorkee), India. \\
\emph{\{shashank\_s1;abhinav\_k1;kris\_k;shivam\_b;pareek\}@ee.iitr.ac.in}}

\maketitle

\begin{abstract}
Generating large-scale, physically consistent AC Optimal Power Flow (ACOPF) datasets is essential for modern data-driven power system applications. The central challenge lies in balancing solution accuracy with computational efficiency. Recent diffusion-based generative models produce high-quality samples; however, their slow sampling procedures limit practical scalability. In this work, we argue that exact physical feasibility is ultimately enforced by power flow solvers or projection steps, and therefore the generative model only needs to produce good initializations rather than perfectly feasible solutions. Based on this insight, we propose a fast diffusion framework using Denoising Diffusion Implicit Models (DDIM) combined with physics-guided corrections during sampling. The proposed method replaces slow stochastic refinement with a small number of deterministic steps and explicit constraint guidance. Experiments on IEEE 6-, 24-, and 118-bus systems show that our approach achieves up to $20\times$ faster sampling than standard diffusion models while maintaining comparable statistical accuracy and physical consistency. This makes the method well suited for scalable OPF dataset generation and practical power system learning tasks. We release the implementation code at \url{https://github.com/PSquare-Lab/DDIM_OPF}.
\end{abstract}




\blfootnote{\noindent $^\star$These authors contributed equally. The order of author names was determined by a game of Ludo. $\dagger$Corresponding Author \emph{pareek@ee.iitr.ac.in}.\\
This work originated as term paper for the a Talent Enhancement Course (EET109) at EE, IIT Roorkee. The authors thank Pranad Lakhote for early discussions. This ongoing work is supported by the ANRF PM Early Career Research Grant (ANRF/ECRG/2024/001962/ENS) and the IIT Roorkee Faculty Initiation Grant (IITR/SRIC/1431/FIG-101078).
}

\section{Introduction}
AC Optimal Power Flow (ACOPF) is central to secure and economic power system operation, supporting real-time control, security assessment and economic dispatch~\cite{19, 20}. The rise of deep learning–based proxy solvers has dramatically increased data demands~\cite{9, 10}, requiring large-scale, physics-consistent datasets to generalize across topologies and rare contingencies such as extreme weather or N-k failures for resilience-focused Digital Twins. However, access to real operational data is severely constrained by privacy, security and market regulations~\cite{11, 12}, forcing reliance on synthetic data generated via repeated solutions of non-convex ACOPF problems~\cite{opfdata}. While accurate, this process is computationally expensive at scale, motivating the need for faster generative alternatives. 

The Generative Adversarial Networks (GANs) enable fast inference, but are prone to \textit{mode collapse}, learning only limited subsets of the target distribution~\cite{8, Arjovsky2017}. In ACOPF, feasible solutions are implicitly defined by nonlinear power flow equations and operational constraints, yielding a highly structured, nonconvex and often disconnected feasible set~\cite{Molzahn2019}. As shown in~\cite{Arjovsky2017}, limited overlap between real and generated distributions causes the Jensen–Shannon divergence gradient to vanish, hindering training. This problem is exacerbated by multiple disconnected feasible regions in ACOPF~\cite{Molzahn2019}, leading GANs to collapse onto a narrow set of operating points. Moreover, restoring feasibility requires costly projection procedures such as Newton–Raphson, largely offsetting GANs’ computational advantages~\cite{Bienstock2019}.

To overcome mode collapse in adversarial training, Denoising Diffusion Probabilistic Models (DDPMs) have emerged as a robust alternative~\cite{2}. By optimizing a stable variational objective and learning the score function, DDPMs capture complex, nonconvex data manifolds and provide full support coverage~\cite{1, 13, 14, 15}. In ACOPF, the learned score field encodes the geometry of the feasibility manifold, guiding samples from noise toward physically meaningful operating regions~\cite{1, 2, 5}. However, standard DDPMs require $\approx 1000$ stochastic refinement steps to preserve their theoretical assumptions, resulting in slow inference~\cite{1, 2}. We argue that such stochastic fidelity is not required for ACOPF generation, where feasibility is ultimately enforced via projection or solver-based post-processing\footnote{Projection-based feasibility recovery is common in ML-based ACOPF methods; see \href{https://energy.hosting.acm.org/wiki/index.php/ML_OPF_wiki}{this} survey for more.}. Our hypothesis is that the diffusion model only needs to initialize samples within the basin of attraction of the projection operator, and fast deterministic diffusion (DDIM) combined with projection offers a more efficient alternative to DDPM-based approximate ACOPF generation.


Main contributions of the paper can be summarized as:
\begin{itemize}
    \item \textbf{Fast diffusion for ACOPF synthesis:} We introduce a DDIM-based generative framework for ACOPF data generation that significantly reduces inference cost by leveraging non-Markovian reverse trajectories.
    \item \textbf{Physics-guided constrained sampling:} We embed ACOPF constraints directly into the DDIM sampling process via gradient-based constraint guidance \cite{1}, steering samples toward the feasible OPF manifold.
    \item \textbf{Scalable and accurate dataset generation:} Extensive experiments on IEEE 6-, 24-, and 118-bus systems demonstrate that the proposed method achieves statistical fidelity comparable to DDPMs while delivering up to $20\times$ faster sampling.
\end{itemize}

\section{Problem Formulation}\label{sec:problem}
The ACOPF problem seeks to minimize system operating costs while satisfying nonlinear physical laws and engineering limits. The power flow equations induce nonconvex equality constraints in ACOPF, and the feasible regions are potentially disjoint~\cite{Molzahn2019}. The general ACOPF problem is classified as NP-hard~\cite{20, Bienstock2019}.


To leverage generative models, we represent each ACOPF solution as a unified state vector $\mathbf{x} \in \mathbb{R}^{4n}$, for an $n$-node system, that captures the complete nodal snapshot:
\begin{equation}\label{eq:x}
\mathbf{x} = [\mathbf{P}^\top, \mathbf{Q}^\top, \mathbf{V}^\top, \boldsymbol{\theta}^\top]^\top
\end{equation}
where, $\mathbf{P} = \mathbf{P}_g - \mathbf{P}_d$ and $\mathbf{Q} = \mathbf{Q}_g - \mathbf{Q}_d$ are the net injections, $\mathbf{V}$ is voltage magnitude vector and $\boldsymbol{\theta}$ is voltage angle vector. This high-dimensional representation allows the model to learn the joint distribution of both the boundary conditions (loads) and the optimal system response (voltages and generation).

\subsection{Generative Synthesis Objective}
Given a ground-truth dataset $\mathcal{D} =\{x_i\}_{i=1}^N$ and the empirical distribution $\mathcal{P}_{\texttt{real}}$ consisting of ACOPF solutions, our objective is to learn a generative distribution $\mathcal{P}_{\texttt{syn}}$ that closely matches the $\mathcal{P}_{\texttt{real}}$ while ensuring all synthesized samples strictly satisfy the physical manifold. This is formulated as:
\begin{subequations}\label{eq:objective}
\begin{align}
\min \quad & \text{dist}(\mathcal{P}_{\texttt{syn}}, \mathcal{P}_{\texttt{real}}), \\
\text{s.t.} \quad & \mathcal{H}(x) = 0 \quad \text{(Power Flow Equations)}, \\
& \mathcal{G}(x) \le 0 \quad \text{(Operational Limits)}.
\end{align}
\end{subequations}
The objective is defined for distribution matching while the equality constraints $\mathcal{H}(x)=0$ enforce the fundamental AC power balance laws:
\begin{subequations}
\begin{align}
P_{i} - |V_i| \sum_{j \in \mathcal{N}} |V_j| (G_{ij} \cos \theta_{ij} + B_{ij} \sin \theta_{ij}) &= 0, \\
Q_{i} - |V_i| \sum_{j \in \mathcal{N}} |V_j| (G_{ij} \sin \theta_{ij} - B_{ij} \cos \theta_{ij}) &= 0,
\end{align}
\end{subequations}
where $\theta_{ij} = \theta_i - \theta_j$ and $G_{ij}, B_{ij}$ are the real and imaginary parts of the bus admittance matrix $Y_{bus}$.
The inequality constraints $\mathcal{G}(x) \le 0$ ensure that the synthesized samples respect the generator capability curves, voltage magnitude regulations and line thermal limits \cite{opfdata}.

Simultaneously achieving distribution matching and constraint satisfaction is non-trivial because feasible ACOPF solutions lie on a highly non-linear, lower-dimensional manifold embedded in $\mathbb{R}^{4n}$. Standard generative models focus primarily on statistical similarity and lack explicit awareness of physical laws, often producing samples that are statistically plausible but physically infeasible. As in \cite{1}, this work addresses this gap by incorporating these physical constraints directly into the diffusion sampling process, enabling the synthesis of diverse, high-fidelity and physically consistent ACOPF datasets.

\section{Diffusion Model Preliminaries}
Diffusion models form the generative backbone of this work and are used to synthesize high-dimensional ACOPF operating points that respect complex physical and operational constraints. At a high level, diffusion-based generative models learn to transform simple noise distributions into structured data distributions through a sequence of denoising steps~\cite{2, 3}. This section briefly reviews DDPMs and motivates the use of DDIMs for efficient dataset generation. 

DDPMs generate data by learning to reverse a gradual noising process. Specifically, a forward diffusion process progressively corrupts clean data samples by adding small amounts of Gaussian noise over a sequence of timesteps. This process is defined as a Markov chain\footnote{In a Markov chain, the next state depends only on the current state and is independent of all earlier states.}
\begin{equation}
q(x_t\!\mid\!x_{t-1})=\mathcal{N}(\sqrt{1-\beta_t}\,x_{t-1},\beta_t I)
\label{eq:forward_process}
\end{equation}
where, ${\beta_t}_{t=1}^T$ is a predefined noise schedule and $x_0$ denotes a clean data sample drawn from the true data distribution.

The learning objective is to train a neural network to approximate the reverse process, which removes noise step-by-step and reconstructs a sample from pure Gaussian noise. During generation the model samples sequentially from $p_\theta(x_{t-1} \mid x_t)$, where the neural network predicts the noise component added at each timestep \cite{2}.

\subsection{The Computational Bottleneck of DDPM}
While Denoising Diffusion Probabilistic Models (DDPMs) exhibit exceptional distributional fidelity~\cite{1}, they suffer from a significant computational bottleneck arising from the inherently stochastic and Markovian nature of the reverse process. Because the transition $p_{\theta}(x_{t-1}|x_t)$ relies on the additive injection of Gaussian noise at every step to approximate Langevin dynamics, the following constraints arise:

\begin{itemize}
    \item \textbf{Small-Step Gaussian Assumption:} The mathematical validity of the reverse transition kernel depends on the assumption that the step size is infinitesimally small. Large steps violate the Gaussian approximation, leading to sampling instability and requiring a large number of fine-grained denoising steps.
    
    \item \textbf{Stochastic Trajectory Redundancy:} The iterative injection of noise transforms the generative path into a high-dimensional random walk. This necessitates $T \approx 1000$ steps to effectively "filter" the stochasticity and converge to the high-density regions of the ACOPF feasibility manifold.
    
    \item \textbf{Discretization Drift:} In a standard Markovian framework, aggressively skipping steps to accelerate inference introduces significant discretization errors. This cumulative variance causes the generated samples to drift away from the valid manifold, resulting in operating points that violate the underlying physical laws~\cite{3}.
\end{itemize}

As a result of this strict step-by-step trajectory, DDPM-based generation is computationally expensive and becomes the primary bottleneck for large scale dataset synthesis, particularly in high-dimensional settings like ACOPF where thousands of samples are required. This limitation motivates the use of a faster sampling scheme: DDIMs.

\section{Proposed Constrained DDIM for ACOPF}

In this section, we describe how DDIM is adapted for generating ACOPF solutions and how physical constraints are incorporated into the diffusion process. Our goal is to generate complete ACOPF states that satisfy network physics and operational limits, while retaining the efficiency and determinism of DDIM sampling. We adopt the unified state representation $\mathbf{x} \in \mathbb{R}^{4n}$ defined in \eqref{eq:x} and avoid any variable decoupling. The overall procedure consists of three components: training the diffusion model on feasible ACOPF data, deterministic DDIM based sampling and constraint guidance to enforce AC power flow feasibility during generation.

\subsection{Training Procedure}
We train a diffusion model to learn the joint distribution of feasible ACOPF solutions, with each training sample given by the state vector $\mathbf{x}$.
This joint representation enables the model to capture the strong physical coupling between power injections and voltages inherent to ACOPF\footnote{Decoupled formulations such as FDLF are intended to accelerate the iterative solution of power-flow equations via numerical approximations. In contrast, our objective is to learn the distribution of feasible ACOPF states, which requires joint modeling to preserve the nonlinear dependencies among $P,Q,V,$ and $\theta$.}. All features are normalized to $[-1,1]$ using min--max scaling computed from the training set. Training follows the standard DDPM forward diffusion process~\cite{ho2020ddpm}, which directly supports DDIM-based inference~\cite{song2020ddim}. During training, a timestep $t \sim \mathcal{U}\{0,\ldots,T-1\}$ is sampled uniformly and noise is added according to
\begin{align}
\mathbf{x}_t
=
\sqrt{\bar{\alpha}_t}\,\mathbf{x}_0
+
\sqrt{1-\bar{\alpha}_t}\,\boldsymbol{\varepsilon},
\qquad
\boldsymbol{\varepsilon} \sim \mathcal{N}(\mathbf{0},\mathbf{I}),
\end{align}
with $\bar{\alpha}_t$ defined by a linear noise schedule~\cite{nichol2021improved}.

The denoising network is a multilayer perceptron (MLP) trained to predict the injected noise using an $\ell_2$ loss, which corresponds to learning the reverse diffusion process and enables fast deterministic DDIM sampling at test time. No physical constraints are imposed during training; the model is trained solely on feasible ACOPF solutions. Feasibility is enforced during inference via physics-guided modifications of the reverse process, as described next.

\subsection{Sampling Procedure}
After training, new ACOPF operating points are generated via a fast reverse diffusion process based on DDIM. Sampling starts from an isotropic Gaussian latent
\begin{equation}
\mathbf{x}_T \sim \mathcal{N}(\mathbf{0},\mathbf{I}),
\end{equation}
representing an unstructured operating point in the normalized ACOPF state space.

At each reverse timestep $t$, the denoising network predicts the noise component $\boldsymbol{\varepsilon}_\theta(\mathbf{x}_t,t)$, which is used to reconstruct a clean estimate of the underlying ACOPF state:
\begin{equation}
\hat{\mathbf{x}}_{0|t}
=
\frac{\mathbf{x}_t - \sqrt{1-\bar{\alpha}_t}\,\boldsymbol{\varepsilon}_\theta(\mathbf{x}_t,t)}
{\sqrt{\bar{\alpha}_t}}.
\end{equation}
Rather than enforcing a strictly stochastic, step-by-step reverse process, we adopt a non-Markovian DDIM update that permits skipping intermediate timesteps while preserving the learned data marginals. This is particularly well suited for ACOPF generation, where the diffusion model only needs to initialize samples within the basin of attraction of a feasibility projection. The resulting transition is
\begin{equation}
\mathbf{x}_{t-1} =
\sqrt{\bar{\alpha}_{t-1}}\,\hat{\mathbf{x}}_{0|t}
+
\sqrt{1 - \bar{\alpha}_{t-1} - \sigma_t^2}\,
\boldsymbol{\varepsilon}_\theta(\mathbf{x}_t,t)
+
\sigma_t \mathbf{z},
\end{equation}
with $\mathbf{z} \sim \mathcal{N}(\mathbf{0},\mathbf{I})$, where $\hat{\mathbf{x}}_{0|t}$ provides the primary denoising direction,
$\boldsymbol{\varepsilon}_\theta(\mathbf{x}_t,t)$ adjusts the reverse
transition to maintain consistency with the learned noise prediction
and $\sigma_t \mathbf{z}$ controls the level of stochasticity.

To enforce physical consistency, the reconstructed estimate $\hat{\mathbf{x}}_{0|t}$ is temporarily mapped to the physical ACOPF variable space, where a differentiable physics-guided correction is applied. This correction balances feasibility and sample diversity along the denoising trajectory. The corrected estimate is then mapped back to the normalized space and used in the DDIM update. Iterating this process over a reduced number of timesteps yields a near-feasible ACOPF solution, which is finally denormalized to obtain the physical operating point.

\subsection{Constrained Guidance for DDIM}

While DDIM enables fast sampling, the generated samples are not guaranteed to satisfy AC power flow physics, since the diffusion model is trained only to match the joint distribution of feasible solutions and does not explicitly enforce Kirchhoff’s laws or operational constraints. As a result, samples obtained purely from the generative process may be statistically plausible yet physically infeasible.

To address this, we incorporate a \emph{constrained guidance} mechanism into the DDIM reverse process~\cite{1}. The key idea is to inject physical knowledge directly into the sampling trajectory by correcting intermediate clean predictions using ACOPF constraints, rather than enforcing feasibility only as a post-processing step. This allows the diffusion trajectory to progressively approach the feasible OPF manifold.

\begin{figure*}[t]
    \centering
    \includegraphics[width=\textwidth]{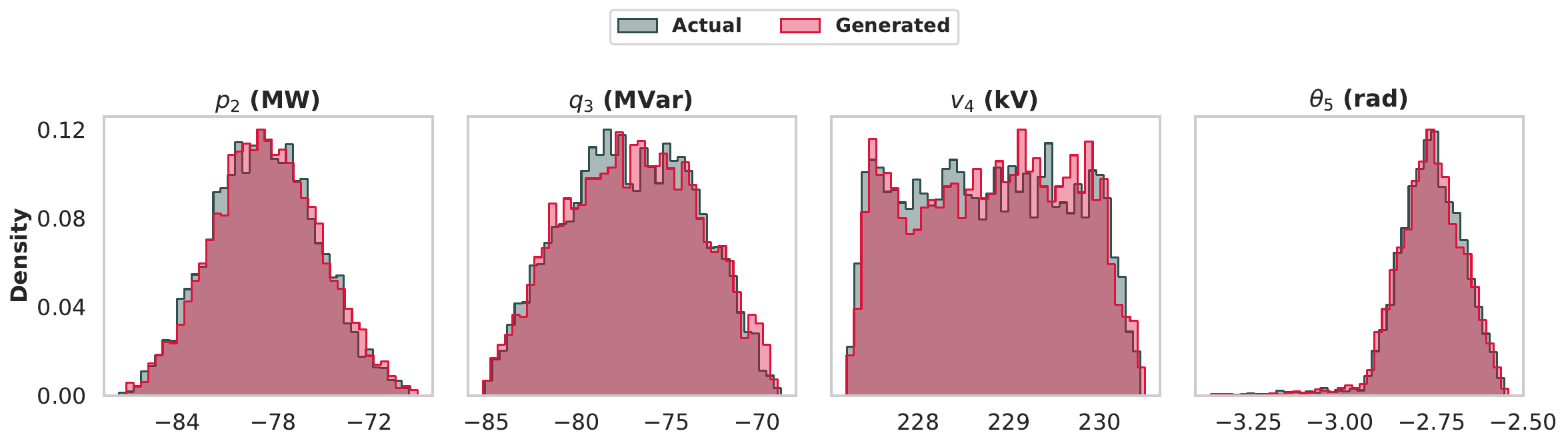}
    \caption{Comparison of ground-truth and generated marginal distributions for selected variables (active power $P$, reactive power $Q$, voltage magnitude $V$ and phase angle $\theta$) in the IEEE 6-bus system.}
    \label{fig:histogram_selected}
\end{figure*}

Using the constraints $H(x)$ and $G(x)$ defined in \eqref{eq:objective}, we introduce smooth residual penalties for both types of constraints as
\begin{align}
R_H(x) = \|H(x)\|_2^2,
\end{align}
\begin{align}
R_G(x) = \|\max(G(x),0)\|_2^2,
\end{align}
which together provide a differentiable measure of physical infeasibility. At each reverse timestep $t$, the DDIM sampler first reconstructs a clean estimate $\hat{\mathbf{x}}_{0|t}$. This estimate is then corrected by taking a gradient step that reduces constraint violations:
\begin{equation}
\hat{\mathbf{x}}'_{0|t}=
\hat{\mathbf{x}}_{0|t}-\lambda_t\nabla_{\mathbf{x}}
\Big(R_H(\hat{\mathbf{x}}_{0|t})+R_G(\hat{\mathbf{x}}_{0|t})\Big),
\end{equation}
where, $\lambda_t$ is a timestep-dependent guidance strength. Stronger corrections are applied at early, high-noise timesteps and gradually relaxed as the sample stabilizes. The corrected estimate $\hat{\mathbf{{x}}}'_{0|t}$ is subsequently used in the DDIM update to compute the next latent state. By injecting constraint information at every reverse step, the generative trajectory is continuously steered toward physically meaningful ACOPF solutions, even when using aggressive timestep skipping.

\begin{algorithm}[H]
\caption{Constrained DDIM Sampling}
\label{alg:ddim_guided}
\begin{algorithmic}[1]
\STATE Initialize $x_T \sim \mathcal{N}(0,I)$ and timestep subset $S$
\FOR{$t \in S$ (descending)}
\STATE $\hat x_0 \leftarrow \frac{x_t-\sqrt{1-\bar\alpha_t}\,\varepsilon_\theta(x_t,t)}{\sqrt{\bar\alpha_t}}$
\STATE $\hat x_0' \leftarrow \hat x_0 - \lambda_t \nabla_x\!\left(R_H(\hat x_0)+R_G(\hat x_0)\right)$
\STATE $x_{t-1} \leftarrow \sqrt{\bar\alpha_{t-1}}\,\hat x_0' + \sqrt{1-\bar\alpha_{t-1}-\sigma_t^2}\,\varepsilon_\theta(x_t,t) + \sigma_t z$
\ENDFOR
\STATE \textbf{Return} $\tilde{x}_0$
\end{algorithmic}
\end{algorithm}

\section{Results}

We evaluate the proposed Constrained DDIM framework on the IEEE 6-, 24-, and 118-bus systems. Performance is assessed along three axes: (i) distributional fidelity, (ii) feature-wise statistical consistency and (iii) sampling efficiency. Visual comparisons (histograms and scatter plots) are used to assess marginal and joint distributions, while quantitative metrics include the 1-Wasserstein distance ($W_1$), KL divergence and wall-clock sampling time.

\subsection{Distributional Fidelity}

\begin{figure*}[t]
    \centering
\includegraphics[width=\textwidth]{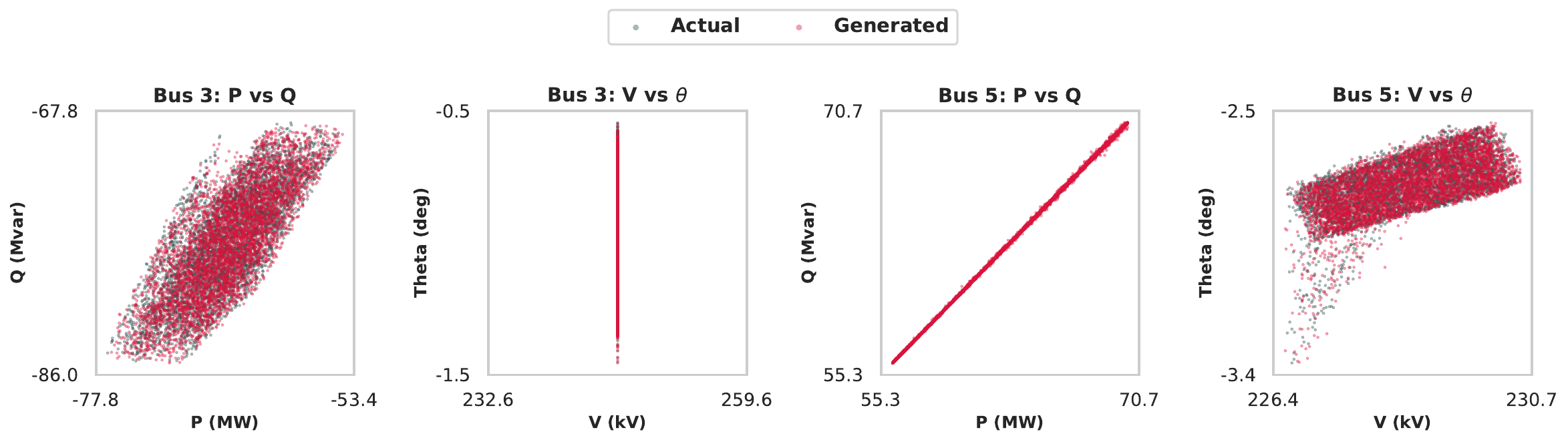}
    \caption{Scatter plots comparing joint distributions of $(P,Q)$ and $(V,\theta)$ for buses 3 and 5 in the IEEE 6-bus system. The strong overlap indicates accurate recovery of inter-variable dependencies.}
    \label{fig:scatter_plots_selected}
\end{figure*}

Fig.~\ref{fig:histogram_selected} compares selected marginal distributions of active power ($P$), reactive power ($Q$), voltage magnitude ($V$) and phase angle ($\theta$) for the IEEE 6-bus system. The generated samples closely match the ground-truth distributions, accurately capturing sharp peaks and boundary effects induced by operational constraints. 
Beyond marginal statistics, Fig.~\ref{fig:scatter_plots_selected} shows bivariate scatter plots for $(P,Q)$ and $(V,\theta)$ pairs. The strong overlap between real and generated samples demonstrates that the proposed method preserves inter-variable dependencies and physical correlations inherent to the AC power flow manifold.

\begin{figure*}[t]
    \centering
    \includegraphics[width=\textwidth]{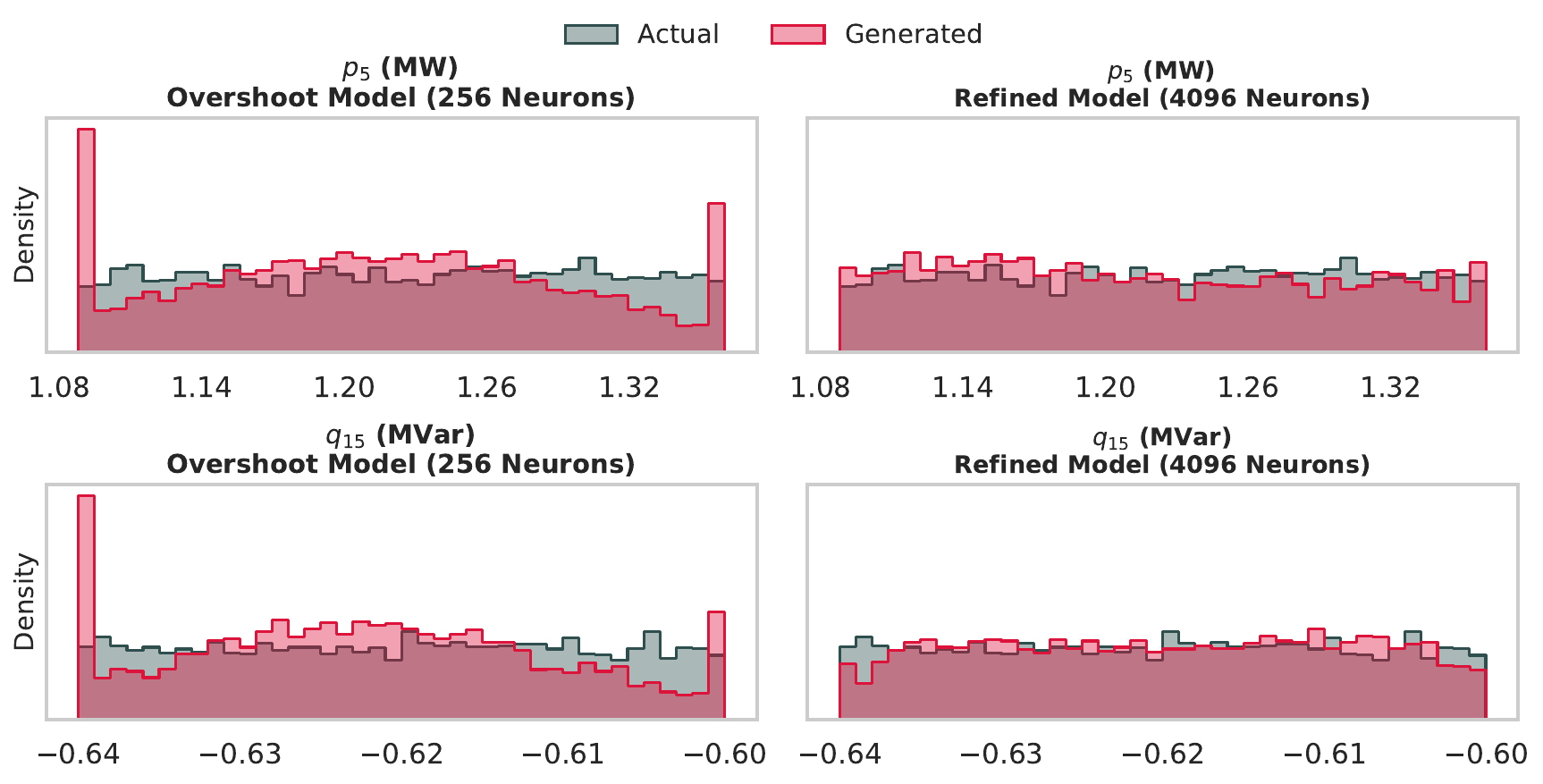}
    \caption{Effect of network capacity on generated marginal distributions for the IEEE 24-bus system. Narrow networks exhibit boundary artifacts, while wider models better capture the empirical distribution.}
    \label{fig:histograms_6bus}
\end{figure*}

\subsection{Statistical Similarity}

Quantitative results are summarized in Table~\ref{tab:combined_res}. Across all systems, DDIM achieves Wasserstein distances comparable to or better than DDPM, despite using significantly fewer reverse diffusion steps. This indicates that accelerated DDIM sampling preserves the global geometry and support of the OPF solution space. Feature-wise fidelity is assessed using KL divergence. DDIM consistently matches or improves upon DDPM across systems, confirming accurate recovery of marginal distributions. Elevated KL values were observed for specific buses and are attributable to zero-injection buses, where even minimal stochastic spread is heavily penalized by KL divergence. This behavior reflects a known limitation of the metric rather than model deficiency. Inorder to retain the true KL divergence value, we explicitly handled these buses during sampling process by clamping their values to zero at every timestep.


\begin{table}[t]
\centering
\caption{Statistical similarity and sampling time comparison for DDPM and DDIM (5,000 samples, $\eta=0.2$, 30 DDIM steps).}
\label{tab:combined_res}
\setlength{\tabcolsep}{4pt}
\renewcommand{\arraystretch}{1.15}
\begin{tabular}{|c|cc|cc|ccc|}
\hline
\multirow{2}{*}{System} 
& \multicolumn{2}{c|}{$W_1$} 
& \multicolumn{2}{c|}{KL} 
& \multicolumn{3}{c|}{Time (s)} \\ \cline{2-8}
& DDPM & DDIM & DDPM & DDIM & OPF & DDPM & DDIM \\ \hline
6-bus   
& 0.028 & 0.018 & 0.027 & 0.023 & 1633 & 57  & 2  \\
24-bus  
& 0.293 & 0.296 & 0.020 & 0.022 & 2637 & 60  & 3  \\
118-bus 
& 0.298 & 0.301 & 0.046 & 0.048 & 4423 & 113 & 6  \\
\hline
\end{tabular}
\end{table}

\subsection{Computational Efficiency}

\begin{figure}[t]
    \centering
    \includegraphics[width=1.0\columnwidth]{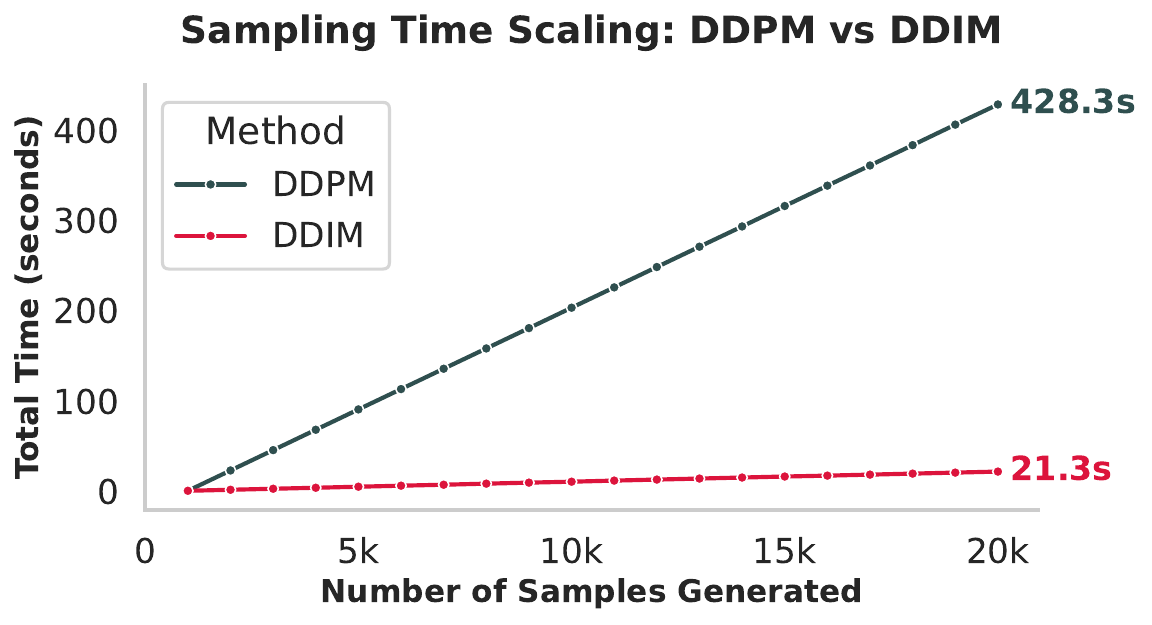}
    \caption{Sampling time versus number of generated samples for the IEEE 118-bus system. DDIM ($\eta=0.2$ and $\text{DDIM steps} = 30$
) achieves approximately $20\times$ speedup over DDPM due to reduced reverse diffusion steps.}
    \label{fig:time_vs_samples}
\end{figure}

Sampling speed is the primary motivation for adopting DDIM. Fig.~\ref{fig:time_vs_samples} shows wall-clock sampling time for up to 20,000 samples on an NVIDIA P100 GPU. DDIM achieves approximately a \textbf{20$\times$ speedup} over DDPM while maintaining comparable statistical fidelity.

Table~\ref{tab:combined_res} further highlights the efficiency gap relative to traditional ACOPF solvers. For the IEEE 118-bus system, DDIM generates 5,000 samples in seconds, compared to several thousand seconds required by a full ACOPF solver. Both DDPM and DDIM exhibit linear scaling with batch size, with DDIM providing a constant-factor acceleration due to reduced reverse steps.


\subsection{Effect of Network Capacity}
\begin{figure}[t]
    \centering
    \includegraphics[width=1.0\columnwidth]{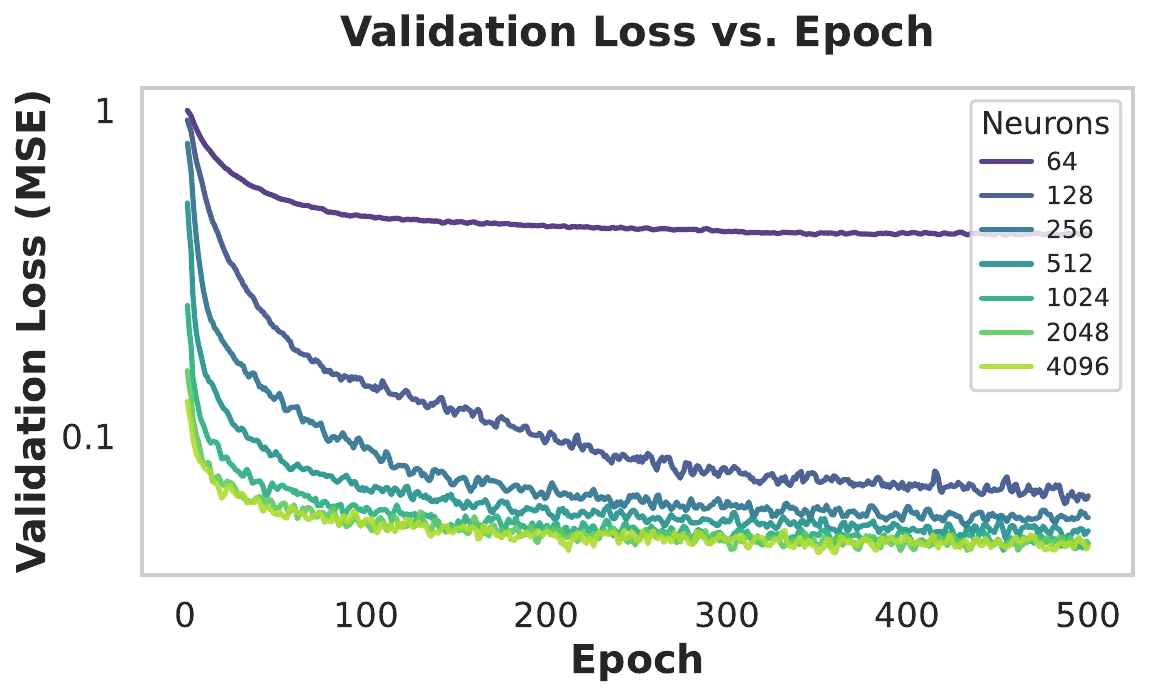}
    \caption{Validation loss versus training epochs for different network widths. Larger models converge faster, with diminishing returns beyond 4096 neurons per layer.}
    \label{fig:validation_vs_epoch}
\end{figure}

Model capacity significantly affects training stability and sample quality. As shown in Fig.~\ref{fig:validation_vs_epoch}, wider networks converge faster and more stably, with performance saturating around 4096 neurons per hidden layer. Lower-capacity models exhibit slower convergence and increased variance. Distributional effects are illustrated in Fig.~\ref{fig:histograms_6bus}. Narrow networks produce visible distortions near operational bounds, while higher-capacity models eliminate these artifacts and better match empirical distributions, indicating improved representation of the nonlinear OPF manifold.

\subsection{Effect of DDIM Stochasticity}

The DDIM stochasticity parameter $\eta$ controls the trade-off between diversity and determinism in the sampling process. Empirically, smaller systems ($n < 30$) perform best with $\eta \approx 0.2$, while larger systems benefit from increased stochasticity, with $\eta \in [0.3, 0.4]$ for $n > 100$. This trend aligns with the increasing dimensionality of the ACOPF state space, which requires additional stochastic exploration to maintain coverage.

\section{Conclusion}
This paper proposed a physics-guided diffusion framework for ACOPF data synthesis using DDIM inference, which learns the joint distribution of $\{P,Q, V, \theta\}$. Experiments on standard IEEE benchmark systems show that DDIM-based sampling, augmented with physics-guided constraint enforcement, achieves up to $20\times$ faster generation than DDPM while maintaining comparable distributional fidelity and physical consistency. These results demonstrate that sampling trajectory design, rather than model complexity, is the dominant factor governing efficiency and feasibility in OPF data generation. Future work will extend the framework to topology-changing scenarios and contingency analysis, incorporate uncertainty-aware constraints and explore integration with real-time OPF solvers and digital twin applications.

\bibliographystyle{IEEEtran}
\bibliography{main}

\end{document}